\documentclass[reprint,superscriptaddress]{revtex4-1}
\usepackage{geometry}                % See geometry.pdf to learn the layout options. There are lots.
\usepackage{graphicx}
\usepackage{amsmath}
\usepackage{amssymb}
\usepackage{epstopdf}
\usepackage{enumitem}
\usepackage{hyperref}
\hypersetup{
    colorlinks,
    citecolor=black,
    filecolor=black,
    linkcolor=red,
    urlcolor=red
}

\begin{document}

\title{Thermodynamic limits for optomechanical systems with conservative potentials}
\date{\today}
\author{Stephen Ragole}
\email{ragole@umd.edu}
\affiliation{Joint Quantum Institute, University of Maryland, \\ College Park, MD 20742, USA}
\affiliation{Joint Center for Quantum Information and Computer Science, University of Maryland, \\ College Park, MD 20742, USA}
\author{Haitan Xu}
\affiliation{Department of Physics \\ Yale University \\ New Haven, CT, 06520, USA}
\author{John Lawall}
\affiliation{National Institute of Standards and Technology \\ Gaithersburg, MD, 20899, USA}
\author{Jacob M. Taylor}
\affiliation{Joint Quantum Institute, University of Maryland, \\ College Park, MD 20742, USA}
\affiliation{Joint Center for Quantum Information and Computer Science, University of Maryland, \\ College Park, MD 20742, USA}
\affiliation{National Institute of Standards and Technology \\ Gaithersburg, MD, 20899, USA}

\begin{abstract}
The mechanical force from light -- radiation pressure -- provides an intrinsic nonlinear interaction. Consequently, optomechanical systems near their steady state, such as the canonical optical spring, can display non-analytic behavior as a function of external parameters. This non-analyticity, a key feature of thermodynamic phase transitions, suggests that there could be an effective thermodynamic description of optomechanical systems. Here we explicitly define the thermodynamic limit for optomechanical systems and derive a set of sufficient constraints on the system parameters as the mechanical system grows large. As an example, we show how these constraints can be satisfied in a system with  $\mathbb{Z}_2$ symmetry and derive a free energy, allowing us to characterize this as an equilibrium phase transition.
\end{abstract}

\maketitle

\section{Introduction}
Phase transitions provide a remarkably powerful framework to study phenomena in many different regimes. While traditionally phase transitions have been studied in classical, equilibrium systems, the most fundamental aspect is the non-analytic behavior of an observable at large system sizes. Looking more generally to non-analytic behavior, others have considered situations that do not meet the strict requirements of thermodynamic equilibrium. In particular, phase transitions have been proposed or observed in systems that are non-equilibrium \cite{Lieb1974}, dissipative \cite{Chakravarty1986}, dynamical \cite{Derrida1987,Fisher1985}, and even quantum mechanical \cite{Sachdev2001,Black2003,Baumann2010,Labeyrie2014,Ritsch2013}. In non-equilibrium systems, numerous analogies with traditional equilibrium phase transitions have been explored, e.g., in lasers \cite{Degiorgio1970,Graham1970}, the Gunn effect \cite{Pytte1968}, and in tunnel diodes \cite{Landauer1961,Landauer1971}. These analogies are fairly broad in consideration, and are readily generalized to other non-equilibrium, non-linear systems, such as those studied in optomechanics. 

In recent years, scientific advances have enabled the creation of numerous optomechanical systems over a range of scales (see  Ref.~\cite{Aspelmeyer2014} for a review). These systems combine the engineerability and control of optical systems with the simplicity of a mechanical harmonic oscillator. Impressive results, including a self-structuring of atoms \cite{Black2003,Baumann2010,Labeyrie2014,Ritsch2013} and a buckling of an optomechanical membrane \cite{Xu2015b} suggest that stable structural rearrangements of the mechanical modes can be described by an order parameter. 

Here we show that the dynamics of the slow modes (the mechanics) can be described by an effective thermodynamic theory despite being an open, non-equilibrium system. Our approach works provided that the fast modes (the optics) obey certain properties, analogous to approaching an optical steady state, defining our thermodynamic limit. This is conceptually similar to integrating out high frequency or short wavelength behavior, but here the process takes a non-equilibrium problem to an equilibrium one, in contrast to the usual formulation of phase transitions. Specifically, we construct a sufficient set of constraints that allow the definition of a thermodynamic limit, and phase transitions, in optomechanical systems. While the limit we define is possible in some cases, we also show a generic optomechanical system may not have such a limit or even be described by a thermodynamic potential. We illustrate our general approach with an example, a phase transition following \cite{Xu2015b}, showing along the way that these constraints are satisfied. While our approach takes into account quantum fluctuations, we do not consider quantum phase transitions in this work. 

\section{Identification of Thermodynamic Limit}

\begin{figure}
	\includegraphics[width=0.45\textwidth]{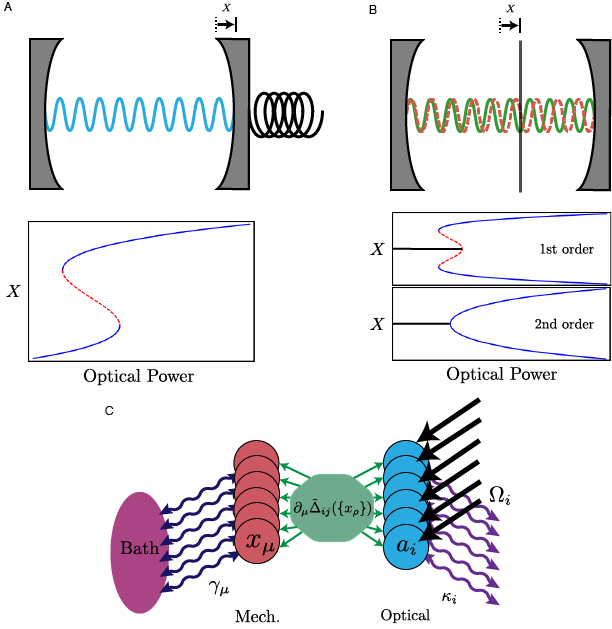}
	\caption{Examples of non-analytic behavior in the steady state mechanics of optomechanical systems. A. Competition between mechanical and optical springs creates bistability as a function of laser power. The stable (unstable) solution is shown as a solid (dashed) line. B. A membrane-in-the-middle system shows a $Z_2$ phase transition which has either a first- (with unstable solutions as dashed lines) or second-order characteristics. C. A cartoon of the generic system with many mechanical ($x_{\mu}$, on the left) and optical modes ($a_i$, on the right) coupled optomechanically and with laser drive ($\Omega_i$) on the optics.}
	\label{fig:fig1}
\end{figure}

We will consider a driven, dissipative system comprising many optical modes coupled to many mechanical modes. Such optomechanical systems have been realized over a wide range of scales, from LIGO to nanoscale resonators or cold atoms \cite{Aspelmeyer2014}. Example non-analytic behavior in these systems is depicted in Figure \ref{fig:fig1}. These systems are, however, far from equilibrium. In the limit that the optics respond instantaneously with respect to the mechanical modes, (i.e., the dynamics for each optical mode are much faster than any of the mechanical frequencies), we may consider the behavior of the optical steady state. Typically, this is accomplished by adiabatically eliminating the optical modes and replacing them with steady state values that depend parametrically on the mechanical modes \cite{Haken1975}. Here we show that we can construct a limit in which the mechanical steady state values are effectively thermodynamic, and identify order parameters in systems with phase transitions.

To ensure that a thermodynamic description holds, we need the system to have conservative dynamics, be stable, have optical forces that are at least as large as mechanical forces and to only couple to bath(s) at a single temperature in the large size limit. Theses constraints (C1-C6) are enumerated in Table \ref{tab:constraints}.
\begin{table}
\caption{The constraints for realizing a thermodynamic limit of an optomechanical system.}
\label{tab:constraints}
\begin{center}
\begin{tabular}{| l | p{6.5cm} |}
	\hline
	$\mathrm{C}1$ & The optical force must have a vanishing curl. \\ \hline
	$\mathrm{C}2$ & The total cavity detuning must remain negative (red-detuned). \\ \hline
	$\mathrm{C}3$ & The optical force must be comparable to the mechanical restoring force. \\ \hline
	$\mathrm{C}4$ & The linearized optical restoring force must remain comparable to the mechanical restoring force. \\ \hline
	$\mathrm{C}5$ & The optically induced damping in the linearized equations must vanish. \\ \hline
	$\mathrm{C}6$ & The optically induced noise in the linearized equations must vanish. \\ \hline
\end{tabular}
\end{center}
\end{table}

We adopt a Hamiltonian formulation for the system, then follow the usual conventions to derive Heisenberg-Langevin equations of motion. As such, we will not specify many of the details of the Hamiltonian, instead focusing on the resulting equations of motion. Still, in principle our system plus bath is described by
\begin{equation}
\small H = H_{\rm opt} + H_{\rm mirror} + H_{\rm mech} + H_{\rm opt bath} + H_{\rm mech bath}\ ,
\end{equation}
where all terms with optical mode operators are assumed to be bilinear or linear in such operators, but may also depend upon the mechanical degrees of freedom. This means that we can write down equations of motion for the optical modes that have no troubles with commutator order, but the addition of optical loss through the mirror into the optical bath will lead to an effective, non-Hermitian picture in the equations of motion approach.

The system obeys the following equations (in the frame rotating with the laser drive frequency for each mode):
\begin{align}
\dot{a}_i &= i\tilde{\Delta}_{ij}(\{x_\rho\})a_j - i \Omega_i+\sqrt{\varkappa^{ex}_{ij}}a^{in}_{j} \\
\dot{x}_{\mu} &= M_{\mu \nu}^{-1} p_{\nu} \label{eq:classical2} \\
\dot{p}_{\mu} &= -K_{\mu \nu} x_{\nu}-\Gamma_{\mu \nu} p_{\nu} +a_i^{\dagger} \partial_{\mu}\tilde{\Delta}_{ij}(\{x_\rho\})a_j \label{eq:classical3} \nonumber \\  &\quad+\sqrt{\Gamma_{\mu \nu}}p^{in}_{\nu} \ , 
\end{align}
where we use Einstein summation notation, with optical modes, $a_i$, indexed by roman indices, coupled to the set of mechanical modes, represented by $x_\mu,\ p_{\mu}$, indexed by greek indices. $\tilde{\Delta}_{ij}(\{x_\rho\})=\Delta_{ij}(\{x_\rho\})+\frac{i}{2}\kappa_{ij}(\{x_\rho\})$ is the non-hermitian matrix, due to $H_{\rm opt} + H_{\rm mirror} + H_{\rm opt bath}$, which describes the dynamics of the optical modes in addition to all of the couplings to the mechanical modes. We note that this generic coupling includes standard dispersive couplings ($i=j$), beam-splitter-like terms ($i\ne j$), and dissipative couplings. $M_{\mu \nu},\ K_{\mu \nu}$, and $\Gamma_{\mu \nu}$ are the matrices, due to $H_{\rm mech} + H_{\rm mech bath}$, giving the effective masses, couplings, and decay rates for the mechanical modes. $\Omega_i$ is the laser drive, $\varkappa^{ex}_{ij}$ is the decay rate matrix for the optical baths, and $a^{in}_j, \ p^{in}_\nu$ are the fluctuations of corresponding optical and mechanical bath fluctuations, respectively. We note that the derivatives of $\tilde{\Delta}_{ij}$ with respect to mechanical coordinates correspond to the vacuum radiation pressure force. Thus, when the optical modes have finite occupation, we will have a finite force.

To separate the steady state effects from the fluctuations, we make the following expansion:
\begin{align}
a_i  &= A_{i} +\delta a_{i} \\
x_\mu  &= X_{\mu} +\delta x_{\mu} \\ 
p_\mu  &= P_{\mu} +\delta p_{\mu}\ ,
 \end{align}
where the capital letters ($A_i,X_\mu,P_\mu$) represent the noiseless, classical variables and the $\delta$ variables are proportional to the fluctuations. We can consider the noiseless variables in the optical steady state ($\dot{A}_i =0$) and study the induced force on the mechanical modes.
\begin{align}
F^{opt}_\mu &= A_{i}^{\dagger}(\{X_\rho\}) \partial_{\mu}\tilde{\Delta}_{ij}(\{X_\rho\})A_{j}(\{X_\rho\})\ .
\end{align}

Our first requirement (C1) is that the curl of this force vanishes, $\epsilon_{\rho \nu \mu}\partial_\nu F_\mu=0$. If this requirement holds, then the mechanics can be described by a conservative potential.
This curl has the form:
\begin{widetext}
\begin{align}
\epsilon_{\rho \nu \mu}\partial_\nu F_\mu &= -\epsilon_{\rho \nu \mu}\bigg(\partial_{\nu} \tilde{\Delta}^\dagger_{il}(\{X_\rho\}) \frac{1}{\tilde{\Delta}_{lm}^{\dagger}(\{X_\rho\})} \partial_{\mu} \tilde{\Delta}_{mj}(\{X_\rho\})\nonumber\\&\qquad \qquad +\partial_{\mu} \tilde{\Delta}_{il}(\{X_\rho\}) \frac{1}{\tilde{\Delta}_{lm}(\{X_\rho\})} \partial_{\nu} \tilde{\Delta}_{mj}(\{X_\rho\})  \bigg)A^{\dagger}_{i}A_{j}\ ,
\end{align}
\end{widetext}
where we have used the fact that partial derivatives commute. There are many possible instances where the curl vanishes. Some example cases are: if there is only a single mechanical mode; a single optical mode; or if $\tilde{\Delta}_{ij}(\{x_\rho\})$ is hermitian - though the lack of damping in the optical modes may violate our adiabatic assumption. We will show a simple case where a non-hermitian $\tilde{\Delta}_{ij}(\{x_\rho\})$ possesses a potential and describes a system with a phase transition. Intriguingly, cases with two or more damped optical modes and multiple mechanical modes generically have a curl, owing to the matrix nature of $\tilde{\Delta}_{ij}(\{x_\rho\})$ and the inclusion of optical loss. Though these cases may have interesting dynamics, including potentially topological properties and limit cycle behaviors, we will not focus on them here.

In the case where the curl vanishes, we need to ensure additional constraints hold to use equilibrium statistical mechanics to describe our system. We need the optical modes to remain red-detuned overall (C2) otherwise instability (via gain) will result. We also require that the optically induced forces remains comparable to the mechanical restoring force in $\{X_\rho\}$ (C3) and in $\{\delta x_\rho\}$ (C4) otherwise the system simply becomes mechanical. While C3-C4 are identical in linear systems, they are distinct in more complicated systems. In stable, open systems considered here, there are (at least) two baths, one optical and one mechanical. For a well-defined (single-temperature) thermodynamic limit, we need the mechanical system to experience a single temperature bath. These restrictions (C5-C6), linked by the fluctuation-dissipation theorem, mean that both the optically-induced mechanical damping and optically-induced mechanical noise must vanish in our limit.

To find the optically-induced forces and the corresponding damping and noise, we study the linearized dynamics of the fluctuations $(\delta a_i, \delta x_{\mu}, \delta p_{\mu})$ which contain the influence from optical and mechanical noise.  To first order, the linearized variables have the following equations of motion:

\begin{align}
\delta\dot{a}_i &= i\tilde{\Delta}_{ij}(\{X_\rho\}) \delta a_j+i\partial_{\mu} \tilde{\Delta}_{ij}(\{X_\rho\})\delta x_{\mu}A_{j} \nonumber \label{eq:linear1} \\ &\quad +\sqrt{\varkappa^{ex}_{ij}}a^{in}_{j}  \\
\delta\dot{x}_{\mu} &= M_{\mu \nu}^{-1} \delta p_{\nu} \label{eq:linear2} \\  \label{eq:linear3}
\delta\dot{p}_{\mu} &= -K_{\mu \nu} \delta x_{\nu} +  A_{i}^{\dagger} \partial_{\mu} \tilde{\Delta}_{ij}(\{X_\rho\}) \delta a_j \nonumber \\ \nonumber &\quad+ \delta a_i^{\dagger}\partial_{\mu} \tilde{\Delta}_{ij}(\{X_\rho\})A_j \\ \nonumber &\quad +A_i^{\dagger}\partial_{\nu} \partial_{\mu} \tilde{\Delta}_{ij}(\{X_\rho\}) \delta x_{\nu} A_j  \\  &\quad- \Gamma_{\mu \nu} \delta p_\nu +\sqrt{\Gamma_{\mu \nu}}p^{in}_{\nu}\ ,
\end{align}
where $a_i^{in},\ p_{\mu}^{in}$ are the bath-induced fluctuations in the optical and mechanical modes respectively.

This is a set of linear equations and can be solved. The solution determines the local stability of the system, and can include both damping and gain, depending upon the sign of the real part of eigenvalues of $\tilde{\Delta}_{ij}(\{X_\rho\})$. We note that this linearized theory is entirely compatible with the full quantum system, but may not capture the full phase diagram of the system outside of our area of focus -- particularly in the regime of limit cycle or oscillator behavior.

We now examine the linear regime in detail. In both the noiseless and linearized equations, we need the optical forces to be comparable to the mechanical forces (C3-C4).  Following our steady state assumption for the optical modes, $A_i=\tilde{\Delta}_{ij}(\{X_\rho\})^{-1}\Omega_j $, we can Fourier transform the equations for the fluctuations, $\delta a_i(t) = \int \mathrm{d}\omega \tilde{a}_i(\omega)e^{-i\omega t}$, expanding in $\omega$:
\begin{widetext}
\begin{align}
 \tilde{a}_i &= \frac{-1}{\omega \delta_{ij}+\tilde{\Delta}_{ij}(\{X_\rho\})}\left(A_{k}\partial_{\mu} \tilde{\Delta}_{jk}(\{X_\rho\})\tilde{x}_\mu  -i\sqrt{\varkappa^{ex}_{jk}}\tilde{a}^{in}_{k}\right) \\ 
&\approx \frac{-1}{\tilde{\Delta}_{ij}(\{X_\rho\})}\left(A_{k}\partial_{\mu} \tilde{\Delta}_{jk}(\{X_\rho\})\tilde{x}_\mu  -i\sqrt{\varkappa^{ex}_{jk}}\tilde{a}^{in}_{k}\right) +\omega\frac{1}{\tilde{\Delta}_{ij}(\{X_\rho\})}\frac{1}{\tilde{\Delta}_{jk}(\{X_\rho\})}\partial_{\mu} \tilde{\Delta}_{kl}(\{X_\rho\})\tilde{x}_\mu A_{l}\ ,
\end{align}
\end{widetext}
where we assume the noise fluctuations are small compared to the optomechanical term $\left(|\sqrt{\varkappa^{ex}_{jk}}\tilde{a}^{in}_{k}| < |A_{k}\partial_{\mu} \tilde{\Delta}_{jk}(\{X_\rho\})\tilde{x}_\mu|\right)$.

\begin{figure}
	\includegraphics[width=0.40\textwidth]{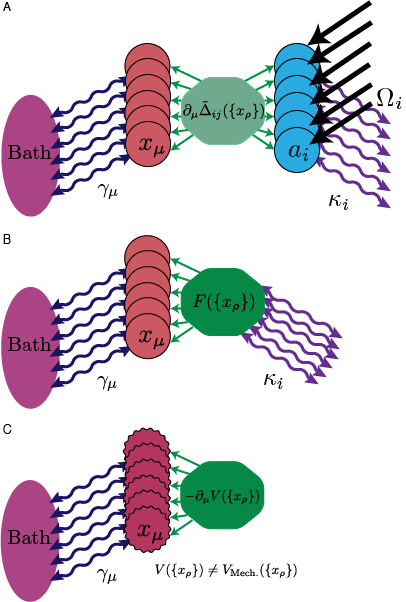}
	\caption{A schematic view of the thermodynamic limit we construct. A. The original system is a generic, driven optomechanical system with an arbitrary number of optical and mechanical modes. B. Using our adiabatic assumption, we use the steady state values for the optical modes, which results in an effective force on the mechanics, including a thermal component from the optical bath. C. Following the constraints, we ensure that the force in B is conservative (C1), stable (C2), comparable to the mechanical force (C3-C4), and that the optical bath fluctuations are negligible (C5-C6), resulting in a mechanical system with a modified potential.}
	\label{fig:schematic}
\end{figure}

Now, inputing the expanded `steady state' optical modes into the Fourier transform of equation \ref{eq:linear3}, we see a coupling to the optical bath which could disrupt the emergence of a thermodynamic limit. These damping and noise effects from $\tilde{a}_i$ must vanish for the mechanics to have a single bath (C5-C6). These six requirements, listed in Table \ref{tab:constraints}, form a set of conditions which must be satisfied as the mechanical modes go to their thermodynamic limit (we envision $X_{\mu} \propto V^\alpha,\ \alpha >0$, where $V$ is the volume of the mechanical resonator and $V\rightarrow \infty$). In Figure \ref{fig:schematic}, we show a simplified view of our approximation and thermodynamic limit process.

If all of these conditions (C1-C6) are met, then the mechanical modes will experience a potential modified by the optics but will not have any additional damping or noise contributions. In this case, the modes will thermalize only to the mechanical reservoir, with no contribution from the optical reservoir. In such a situation, we can compute the partition function and upon integration of the mechanical modes, determine a free energy for an order parameter if one exists in the system.

There are interesting systems which do not meet these conditions, however, we restrict ourselves to the case where our effective thermodynamic theory applies. In particular, we demonstrate the existence of a $\mathbb{Z}_2$ phase transition in the defined thermodynamic limit of an optomechanical membrane-in-the-middle system.

\section{The Thermodynamic Limit for the $\mathbb{Z}_2$ System}

We consider an optomechanical system with two optical modes, $a_{1,2}$, coupled oppositely to a single mechanical mode, $x$, with resonant frequency $\Omega_m$, where each optical mode has equal drive and decay, ($\Omega_{1,2}=\Omega,\ \kappa_{1,2} = \kappa$),. Explicitly, we define $\Delta_{11} =\Delta_1+ gx$ and $\Delta_{22} =\Delta_2-gx$, where $\Delta_1=\Delta_2=\Delta$ is the detuning of the modes when $x=0$. As an example, one can consider a cavity with a dielectric membrane-in-the-middle \cite{Xu2015b,Jayich2008a,Jayich2008c,Sankey2010a} where we drive two modes with opposite responses to the membrane motion, depicted in Fig. \ref{fig:fig1}B. This system has been realized experimentally and shows the characteristics of a phase transition \cite{Xu2015b}, which we expand upon.

\begin{figure}
	\includegraphics[width=.45\textwidth]{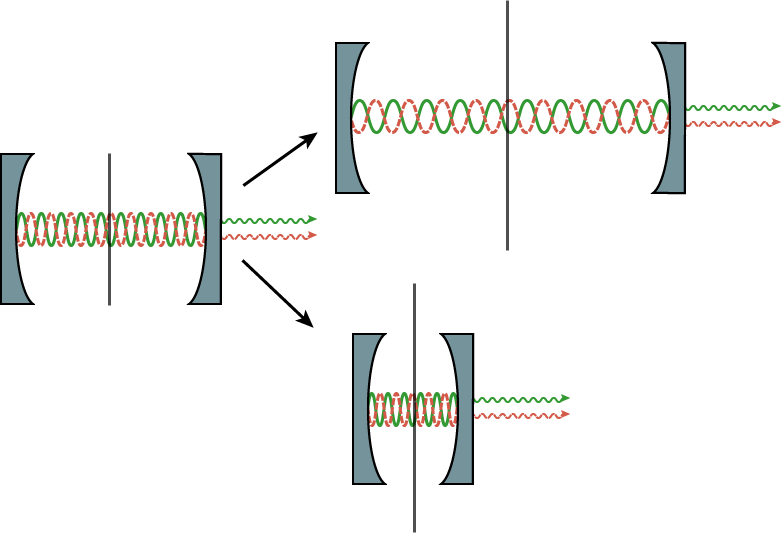}
	\caption{Two possible realization of the thermodynamic limit for the membrane-in-the-middle system. In the upper figure, we imagine the cavity growing with the membrane, while the finesse decreases to ensure $\kappa$ stays large. In the lower figure, we consider shrinking the cavity, which meets our constraints so long as the finesse does not increase more quickly than the cavity shrinks. In each case, the power required to satisfy the other constraints grows.}
	\label{fig:ThermoLimit}
\end{figure}

To demonstrate that such a system has a thermodynamic limit, we need to determine how the above constraints (C1-C6) are held. We choose the `bad cavity limit,' $\kappa_i \gg \Omega_m, \gamma$, such that the optical decay is much faster than any mechanical frequency. C1 is satisfied immediately because the curl of a one-dimensional force vanishes identically. We expand the variables ($a_1, a_2, x, p$) and imagine that $X\propto V^\alpha \rightarrow \infty$, as above. We also consider the scaling of the membrane mass, $m = \rho V$, the coupling, $g = \omega_c d^{-1}$, and the cavity decay, $\kappa = c d^{-1} \mathcal{F}^{-1}$ where $\rho$ is the mass density and $\omega_c, d, \mathcal{F}$ are the cavity frequency, length, and finesse, respectively. We can consider a variety of options for the cavity length, $d$, and adjust the other parameters, such as finesse and cavity frequency, to ensure the following scaling arguments hold. Two options for the cavity scaling are depicted in Figure \ref{fig:ThermoLimit}. With C1 automatically satisfied, the rest of the constraints become a set of scaling relations that determine a region of parameter space in which the defined thermodynamic limit exists.

To ensure the cavity stays red-detuned (C2) we must impose that $\Delta \propto -|gX|$. Taking the steady state solutions for $A_i,\tilde{a}_i$ as described above, we can derive the optically induced force, damping, and noise in order to quantify the remaining constraints. The optical and linearized optical forces (C3-C4) are:
\begin{widetext}
\begin{align}
F^{opt} &= -\frac{4\hbar g^2 \Delta |\Omega|^2}{(\Delta^2+\frac{\kappa^2}{4})^2+2g^2X^2(\frac{\kappa^2}{4}-\Delta^2)+g^4X^4}X\ ,\\
f^{opt} &= -\frac{4\hbar g^2  \Delta |\Omega|^2\left((\Delta^2+\frac{\kappa^2}{4})^2+2g^2X^2(\Delta^2+\frac{\kappa^2}{4})-3g^4X^4\right)}{\left((\Delta^2+\frac{\kappa^2}{4})^2+2g^2X^2(\frac{\kappa^2}{4}-\Delta^2)+g^4X^4\right)^2}\delta x\ ,
\end{align}
\end{widetext}
The optical force must be comparable to $m\Omega_m^2 X$, while the linearized optical force must be comparable to $m\Omega_m^2 \delta x$. Since $m\Omega_m^2$ remains finite in $2D$, the coefficient of these optical springs must not vanish. Given the linear nature of the coupling, C3 and C4 are identical constraints which are satisfied when $|\Omega|^2  \ge c_{3} |gX|^3$ and $\kappa \le c_{4}|gX|$, where $c_3,c_4$ are fixed constants, and we used C2 to achieve this result.

Having established that the force is relevant for the steady state position and its fluctuations, we can consider dissipative effects. The damping and noise will be carried into the mechanical equations from $\tilde{a}$. These terms must vanish if we are to achieve the desired single temperature bath (C5-C6).  The damping is:
\begin{widetext}
\begin{align}
 \gamma^{opt} &=  -\frac{4 \hbar g^2 |\Omega|^2 \Delta \kappa\left(5g^6X^6+g^4X^4 (5\Delta^2 + \frac{9\kappa^2}{4})+3g^2X^2(\frac{\kappa^4}{16}-\frac{1}{2} \Delta^2 \kappa^2 - 3 \Delta^4)-(\Delta^2+\frac{\kappa^2}{4})^3\right)}{m\left((\Delta^2+\frac{\kappa^2}{4})^2+2g^2X^2(\frac{\kappa^2}{4}-\Delta^2)+g^4X^4\right)^3}\ ,
\end{align}
\end{widetext}
which, following C1-C4, scales at most as $\frac{\Omega_m^2}{|gX|}$ and thus vanishes as $X \rightarrow \infty$, satisfying C5.

Finally, we consider the optically induced noise (C6).  These noise terms have the form:
\begin{align}
b_{in_i} = \frac{\hbar g |\Omega| \sqrt{\kappa} a_{i,in}}{\left((\Delta \pm gX) + \frac{i\kappa}{2}\right)^2}\ ,
\end{align}
where $b_{in_i}$ is the noise term in $\dot{p}$ and $i=1,2$ determines the sign of the $gx$.

These terms should be considered in relation to the mechanical noise, i.e., we should compute $\frac{\langle b^{\dagger}_{in}b_{in}\rangle}{\langle p^{\dagger}_{in} p_{in} \rangle}$ where $p_{in}$ is the noise from the mechanical bath. Assuming the mechanics have an ohmic bath, this ratio is:
\begin{align}
\frac{\langle b^{\dagger}_{in}b_{in}\rangle}{\langle p^{\dagger}_{in} p_{in} \rangle} &= \frac{\hbar^2 g^2 |\Omega|^2 \kappa}{\left((\Delta \pm gX)^2+\frac{\kappa^2}{4}\right)^2}\frac{1}{2m\gamma k_bT}\ ,
\end{align}
where the plus (minus) corresponds to $b_{1(2)}$.

Following the scaling from above, this term is proportional to $\frac{\hbar \Omega_m^2 \kappa}{2 gX \gamma k_b T}= \frac{Q_{\textrm{mq}} \kappa}{gX}$, where $Q_{\textrm{mq}} = \frac{\hbar \Omega_m^2}{\gamma k_b T}$ is the quantum $Q$ for the oscillator. From above, we need $\kappa \le c_4 |gX|$ but we also need to consider how $Q_{\textrm{mq}}$ behaves in the thermodynamic limit. From the scaling of $X$, we have $Q_{\textrm{mq}} \propto  \frac{1}{V^{2\alpha} \gamma}$ which vanishes so long as $\gamma > c_5 V^{-2\alpha}$, i.e., if the mechanical noise stays finite, it will overwhelm the optically induced noise and form the dominant noise contribution.

With this final constraint in place, we have demonstrated that our optomechanical system has a well defined potential with only relevant coupling to a single bath in the thermodynamic limit, which can be described by equilibrium statistical mechanics.  

\section{The Free Energy of the $\mathbb{Z}_2$ System}
We generalize this analysis to include many mechanical modes, such as those present in a membrane, each coupled to the optics in the same fashion (though, potentially with different values of $g$). We compute the partition function for these membrane modes and the optically induced potential. We will consider a membrane with the displacement field  $u(\vec{r},t)$, conjugate momentum $\pi(\vec{r},t)$, a mass density $\rho$, and a Young's modulus, $Y$. As an order parameter, we identify $O(t) = \int \mathrm{d}^2r g(\vec{r}) u(\vec{r},t)$, which appears in the optical potential, $V(O)$. We can write the classical membrane Hamiltonian and the full partition function including the optical potential:
\begin{align}
H &= \int \mathrm{d}^2 r \big[\frac{ \pi(\vec{r},t)^2}{2\rho}+Y\left(\partial_\mu u(\vec{r},t) \right)^2 \nonumber \\ &\qquad +f(\vec{r},t)u(\vec{r},t) \big] \\
Z&= \int \mathcal{D}u\mathcal{D}\pi \mathrm{d}O \mathrm{d}\lambda  [e^{-\beta (H[u,\pi]+V(O))} \nonumber \\ &\qquad\qquad \times e^{i\lambda (O - \int \mathrm{d}^d r g(\vec{r}) u(\vec{r},t))}]\ ,
\end{align}
where $f(\vec{r},t)$ is an external force that might break the symmetry, and we have added the order parameter, $O$, in as an auxiliary field. 

After transforming to Fourier space and integrating out the membrane momentum, $\pi$, displacement, $u$, and $\lambda$, we can rewrite the partition function (up to normalization factors) purely in terms of the order parameter $O$, renaming the effective `spring constant,' $k=\left(\sum_{\vec{k}} \frac{g_{-k}^2}{\Omega_{m_k}^2}\right)^{-1}$:
\begin{align}
Z' &= \int \mathrm{d}O \big[ \exp\left(-\beta\left( \frac{k}{2}\left(O-\sum_k\frac{g_{k}f_{k}}{\Omega_{m_k}^2}\right)^2 \right)\right)\nonumber \\ & \qquad \times \exp{\left(-\beta V(O)\right)}\big]\ , \label{eqn:partition}
\end{align}
where $g_k,f_k$ are the Fourier components of the optomechanical coupling and external forces respectively. From Eqn. \ref{eqn:partition}, we can perform the integral and compute the free energy, F:
\begin{align}
F &= \frac{1}{\beta} \log{Z'}\ .
\end{align}
Since $O$ grows with system size (and the potential along with it), only the minima of the effective potential will have finite energy in the thermodynamic limit. These correspond to the usual saddle points in steepest descent approximations.

The locations of the minima of $F$ change non-analytically as a function of laser power, resulting in a phase transition in the steady state of the coupled mode. Intriguingly, the order and onset of the phase transition are strongly dependent on the detuning of the optical modes from the cavity. To see this behavior, we study the values of the order parameter $O$ which give zero `force'.  We compute the zero force condition: \small
\begin{align}
0 &= k O \left(1 - \frac{A}{O^4+2O^2(\frac{\kappa^2}{4}-\Delta^2)+(\Delta^2+\frac{\kappa^2}{4})} \right)\ ,
\end{align} \normalsize
where we define $A = \frac{-4\hbar g^2 |\Omega|^2 \Delta}{k}$, where $\Omega$ is the coupling to the laser drive. This quantity is positive since we are considering red-detuning ($\Delta<0$) and is proportional to the input laser power. We also note that the second term has even parity, so solutions with $O\ne 0$ will appear in pairs, providing the $\mathbb{Z}_2$ symmetry of the steady state solutions anticipated.

The general solution for the non-trivial phase  ($O \ne 0$) is
\begin{align}
\label{eqn:soln}
O^2 &= \Delta^2 -\frac{\kappa^2}{4} \pm \sqrt{A-\Delta^2\kappa^2}\ .
\end{align}
For the solution to be valid, the right hand side needs to be real and positive. Enforcing realness, $A > \Delta^2 \kappa^2$. Recalling that $A$ is proportional to laser power, this constraint gives the minimum laser power for a transition to occur.
To ensure positivity, we have to consider two cases, $|\Delta| < \frac{\kappa}{2}$ and  $|\Delta| > \frac{\kappa}{2}$. In the first case, the square root term must be larger than the first two terms, so only the positive root can be a solution. For that to be the case, we need
\begin{align}
A-\Delta^2\kappa^2 &>\left(\frac{\kappa^2}{4}-\Delta^2\right)^2 \nonumber \\ 
A &> \left(\frac{\kappa^2}{4}-\Delta^2\right)^2+\Delta^2\kappa^2 \nonumber \\
A &> \left(\frac{\kappa^2}{4}+\Delta^2\right)^2\ .
\end{align}
When this equation is satisfied, there is only one solution for $O^2$. This case corresponds to a double well potential where the wells split from $O=0$ as power is increased. The $O=0$ solution becomes unstable and represents the peak of the barrier between the two wells at $O = \pm O_s$.  However, in the case where $|\Delta| > \frac{\kappa}{2}$, once $A > \Delta^2 \kappa^2$, a triple well develops with minima at $O=0,\pm O_{s+}$, where $O_{s+}$ is the larger solution. We will show below that the smaller solution, $O_{s-}$, is unstable and forms the peaks of the barriers.

To study the stability of the solutions, necessary for the steepest descents approximation, we determine the local curvature of the potential at each of these critical points by computing the second derivative of the potential, $\partial_{O}^2 V(O)$. Defining $u=O^2$ and $D(u) = u^2+2u(\frac{\kappa^2}{4}-\Delta^2)+(\Delta^2+\frac{\kappa^2}{4})$ for convenience,
 \begin{equation}
\partial_{O}^2 V(O) = k\left[1 - \frac{A}{D(u)} \right]+ \frac{kO A}{D(u)^2}D'(u)2O\ .
\end{equation}
The $O_{s}=0$ solution is stable but decreasingly so until $\frac{A}{D(0)}=1$ (which is the power at which as the $|\Delta| <\frac{\kappa}{2}$ case buckles) and the solution becomes unstable.

In the buckled state, we see the mechanical spring constant drops out immediately and the curvature is controlled by the optical response. The sign of the curvature is determined by $ D'(O_{s\pm}^2)= \pm 2\sqrt{A-\Delta^2\kappa^2}$.
Therefore, when $|\Delta| > \frac{\kappa}{2}$, the smaller solutions are unstable, leading to a first order phase transition, while the outer solutions and the solutions for $|\Delta| < \frac{\kappa}{2}$ are stable with a new optically determined spring constant. Thus, as a function of laser power, the steady state of the membrane will either experience a 1st or 2nd order phase transitions which spontaneously breaks the $\mathbb{Z}_2$ symmetry of the potential.

\section{Conclusion}
We have developed a set of constraints on optomechanical systems under which an effective equilibrium thermodynamic phase transition can be defined. These constraints, described in Table~\ref{tab:constraints}, allow the mechanics to explore an optically modified potential in a regime where the effects of optical fluctuations are overwhelmed by mechanical ones such that it equilibrates to the mechanical bath temperature. Exploring this limit, we define a system which possesses a phase transition of either 1st or 2nd order, controlled by the system parameters. 
Specifically, our theory supports the experimental observation of spontaneous $\mathbb{Z}_2$ symmetry breaking corresponding to the buckling of the mechanical spring \cite{Xu2015b}.

We also find, more generally, that optomechanical systems which do not have conservative dynamics are generic, and are not well understood in our thermodynamic limit. Analysis of these systems may point to topological physics and connect with other related optomechanical systems, such as those with exceptional points \cite{Xu2016}. While experimental efforts have included the ability to cool the mechanical system to its ground state \cite{Chan2011}, determining whether our framework for phase transitions persists at the quantum level will require further analysis to handle the effects of quantum fluctuations. However, the possibility of observing quantum phase transitions which spontaneously break symmetry remain quite compelling and will drive future theoretical work.

\bibliographystyle{apsrev4-1}
\bibliography{Z2Paper3}

%merlin.mbs apsrev4-1.bst 2010-07-25 4.21a (PWD, AO, DPC) hacked
%Control: key (0)
%Control: author (72) initials jnrlst
%Control: editor formatted (1) identically to author
%Control: production of article title (-1) disabled
%Control: page (0) single
%Control: year (1) truncated
%Control: production of eprint (0) enabled
\begin{thebibliography}{22}%
\makeatletter
\providecommand \@ifxundefined [1]{%
 \@ifx{#1\undefined}
}%
\providecommand \@ifnum [1]{%
 \ifnum #1\expandafter \@firstoftwo
 \else \expandafter \@secondoftwo
 \fi
}%
\providecommand \@ifx [1]{%
 \ifx #1\expandafter \@firstoftwo
 \else \expandafter \@secondoftwo
 \fi
}%
\providecommand \natexlab [1]{#1}%
\providecommand \enquote  [1]{``#1''}%
\providecommand \bibnamefont  [1]{#1}%
\providecommand \bibfnamefont [1]{#1}%
\providecommand \citenamefont [1]{#1}%
\providecommand \href@noop [0]{\@secondoftwo}%
\providecommand \href [0]{\begingroup \@sanitize@url \@href}%
\providecommand \@href[1]{\@@startlink{#1}\@@href}%
\providecommand \@@href[1]{\endgroup#1\@@endlink}%
\providecommand \@sanitize@url [0]{\catcode `\\12\catcode `\$12\catcode
  `\&12\catcode `\#12\catcode `\^12\catcode `\_12\catcode `\%12\relax}%
\providecommand \@@startlink[1]{}%
\providecommand \@@endlink[0]{}%
\providecommand \url  [0]{\begingroup\@sanitize@url \@url }%
\providecommand \@url [1]{\endgroup\@href {#1}{\urlprefix }}%
\providecommand \urlprefix  [0]{URL }%
\providecommand \Eprint [0]{\href }%
\providecommand \doibase [0]{http://dx.doi.org/}%
\providecommand \selectlanguage [0]{\@gobble}%
\providecommand \bibinfo  [0]{\@secondoftwo}%
\providecommand \bibfield  [0]{\@secondoftwo}%
\providecommand \translation [1]{[#1]}%
\providecommand \BibitemOpen [0]{}%
\providecommand \bibitemStop [0]{}%
\providecommand \bibitemNoStop [0]{.\EOS\space}%
\providecommand \EOS [0]{\spacefactor3000\relax}%
\providecommand \BibitemShut  [1]{\csname bibitem#1\endcsname}%
\let\auto@bib@innerbib\@empty
%</preamble>
\bibitem [{\citenamefont {Lieb}(1974)}]{Lieb1974}%
  \BibitemOpen
  \bibfield  {author} {\bibinfo {author} {\bibfnamefont {E.~H.}\ \bibnamefont
  {Lieb}},\ }\href {\doibase 10.1016/0031-8914(74)90237-7} {\bibfield
  {journal} {\bibinfo  {journal} {Physica}\ }\textbf {\bibinfo {volume} {73}},\
  \bibinfo {pages} {226} (\bibinfo {year} {1974})}\BibitemShut {NoStop}%
\bibitem [{\citenamefont {Chakravarty}\ \emph {et~al.}(1986)\citenamefont
  {Chakravarty}, \citenamefont {Ingold}, \citenamefont {Kivelson},\ and\
  \citenamefont {Luther}}]{Chakravarty1986}%
  \BibitemOpen
  \bibfield  {author} {\bibinfo {author} {\bibfnamefont {S.}~\bibnamefont
  {Chakravarty}}, \bibinfo {author} {\bibfnamefont {G.~L.}\ \bibnamefont
  {Ingold}}, \bibinfo {author} {\bibfnamefont {S.}~\bibnamefont {Kivelson}}, \
  and\ \bibinfo {author} {\bibfnamefont {A.}~\bibnamefont {Luther}},\ }\href
  {\doibase 10.1103/PhysRevLett.56.2303} {\bibfield  {journal} {\bibinfo
  {journal} {Phys. Rev. Lett.}\ }\textbf {\bibinfo {volume} {56}},\ \bibinfo
  {pages} {2303} (\bibinfo {year} {1986})}\BibitemShut {NoStop}%
\bibitem [{\citenamefont {Derrida}(1987)}]{Derrida1987}%
  \BibitemOpen
  \bibfield  {author} {\bibinfo {author} {\bibfnamefont {B.}~\bibnamefont
  {Derrida}},\ }\href {\doibase 10.1088/0305-4470/20/11/009} {\bibfield
  {journal} {\bibinfo  {journal} {J. Phys. A. Math. Gen.}\ }\textbf {\bibinfo
  {volume} {20}},\ \bibinfo {pages} {L721} (\bibinfo {year}
  {1987})}\BibitemShut {NoStop}%
\bibitem [{\citenamefont {Fisher}(1985)}]{Fisher1985}%
  \BibitemOpen
  \bibfield  {author} {\bibinfo {author} {\bibfnamefont {D.~S.}\ \bibnamefont
  {Fisher}},\ }\href {\doibase 10.1103/PhysRevB.31.1396} {\bibfield  {journal}
  {\bibinfo  {journal} {Phys. Rev. B}\ }\textbf {\bibinfo {volume} {31}},\
  \bibinfo {pages} {1396} (\bibinfo {year} {1985})}\BibitemShut {NoStop}%
\bibitem [{\citenamefont {Sachdev}(2011)}]{Sachdev2001}%
  \BibitemOpen
  \bibfield  {author} {\bibinfo {author} {\bibfnamefont {S.}~\bibnamefont
  {Sachdev}},\ }\href@noop {} {\emph {\bibinfo {title} {{Quantum Phase
  Transitions}}}},\ \bibinfo {edition} {2nd}\ ed.\ (\bibinfo  {publisher}
  {Cambridge Univ. Press},\ \bibinfo {year} {2011})\BibitemShut {NoStop}%
\bibitem [{\citenamefont {Black}\ \emph {et~al.}(2003)\citenamefont {Black},
  \citenamefont {Chan},\ and\ \citenamefont {Vuletic}}]{Black2003}%
  \BibitemOpen
  \bibfield  {author} {\bibinfo {author} {\bibfnamefont {A.~T.}\ \bibnamefont
  {Black}}, \bibinfo {author} {\bibfnamefont {H.~W.}\ \bibnamefont {Chan}}, \
  and\ \bibinfo {author} {\bibfnamefont {V.}~\bibnamefont {Vuletic}},\ }\href
  {\doibase 10.1103/PhysRevLett.91.203001} {\bibfield  {journal} {\bibinfo
  {journal} {Phys. Rev. Lett.}\ }\textbf {\bibinfo {volume} {91}},\ \bibinfo
  {pages} {203001/1} (\bibinfo {year} {2003})}\BibitemShut {NoStop}%
\bibitem [{\citenamefont {Baumann}\ \emph {et~al.}(2010)\citenamefont
  {Baumann}, \citenamefont {Guerlin}, \citenamefont {Brennecke},\ and\
  \citenamefont {Esslinger}}]{Baumann2010}%
  \BibitemOpen
  \bibfield  {author} {\bibinfo {author} {\bibfnamefont {K.}~\bibnamefont
  {Baumann}}, \bibinfo {author} {\bibfnamefont {C.}~\bibnamefont {Guerlin}},
  \bibinfo {author} {\bibfnamefont {F.}~\bibnamefont {Brennecke}}, \ and\
  \bibinfo {author} {\bibfnamefont {T.}~\bibnamefont {Esslinger}},\ }\href
  {\doibase 10.1038/nature09009} {\bibfield  {journal} {\bibinfo  {journal}
  {Nature}\ }\textbf {\bibinfo {volume} {464}},\ \bibinfo {pages} {1301}
  (\bibinfo {year} {2010})}\BibitemShut {NoStop}%
\bibitem [{\citenamefont {Labeyrie}\ \emph {et~al.}(2014)\citenamefont
  {Labeyrie}, \citenamefont {Tesio}, \citenamefont {Gomes}, \citenamefont
  {Oppo}, \citenamefont {Firth}, \citenamefont {Robb}, \citenamefont {Arnold},
  \citenamefont {Kaiser},\ and\ \citenamefont {Ackemann}}]{Labeyrie2014}%
  \BibitemOpen
  \bibfield  {author} {\bibinfo {author} {\bibfnamefont {G.}~\bibnamefont
  {Labeyrie}}, \bibinfo {author} {\bibfnamefont {E.}~\bibnamefont {Tesio}},
  \bibinfo {author} {\bibfnamefont {P.~M.}\ \bibnamefont {Gomes}}, \bibinfo
  {author} {\bibfnamefont {G.-L.}\ \bibnamefont {Oppo}}, \bibinfo {author}
  {\bibfnamefont {W.~J.}\ \bibnamefont {Firth}}, \bibinfo {author}
  {\bibfnamefont {G.~R.~M.}\ \bibnamefont {Robb}}, \bibinfo {author}
  {\bibfnamefont {a.~S.}\ \bibnamefont {Arnold}}, \bibinfo {author}
  {\bibfnamefont {R.}~\bibnamefont {Kaiser}}, \ and\ \bibinfo {author}
  {\bibfnamefont {T.}~\bibnamefont {Ackemann}},\ }\href {\doibase
  10.1038/nphoton.2014.52} {\bibfield  {journal} {\bibinfo  {journal} {Nat.
  Photonics}\ }\textbf {\bibinfo {volume} {8}},\ \bibinfo {pages} {321}
  (\bibinfo {year} {2014})},\ \Eprint {http://arxiv.org/abs/1308.1226}
  {arXiv:1308.1226} \BibitemShut {NoStop}%
\bibitem [{\citenamefont {Ritsch}\ \emph {et~al.}(2013)\citenamefont {Ritsch},
  \citenamefont {Domokos}, \citenamefont {Brennecke},\ and\ \citenamefont
  {Esslinger}}]{Ritsch2013}%
  \BibitemOpen
  \bibfield  {author} {\bibinfo {author} {\bibfnamefont {H.}~\bibnamefont
  {Ritsch}}, \bibinfo {author} {\bibfnamefont {P.}~\bibnamefont {Domokos}},
  \bibinfo {author} {\bibfnamefont {F.}~\bibnamefont {Brennecke}}, \ and\
  \bibinfo {author} {\bibfnamefont {T.}~\bibnamefont {Esslinger}},\ }\href
  {\doibase 10.1103/RevModPhys.85.553} {\bibfield  {journal} {\bibinfo
  {journal} {Rev. Mod. Phys.}\ }\textbf {\bibinfo {volume} {85}},\ \bibinfo
  {pages} {553} (\bibinfo {year} {2013})},\ \Eprint
  {http://arxiv.org/abs/1210.0013} {arXiv:1210.0013} \BibitemShut {NoStop}%
\bibitem [{\citenamefont {Degiorgio}\ and\ \citenamefont
  {Scully}(1970)}]{Degiorgio1970}%
  \BibitemOpen
  \bibfield  {author} {\bibinfo {author} {\bibfnamefont {V.}~\bibnamefont
  {Degiorgio}}\ and\ \bibinfo {author} {\bibfnamefont {M.~O.}\ \bibnamefont
  {Scully}},\ }\href {\doibase 10.1103/PhysRevA.2.1170} {\bibfield  {journal}
  {\bibinfo  {journal} {Phys. Rev. A}\ }\textbf {\bibinfo {volume} {2}},\
  \bibinfo {pages} {1170} (\bibinfo {year} {1970})}\BibitemShut {NoStop}%
\bibitem [{\citenamefont {Graham}\ and\ \citenamefont
  {Haken}(1970)}]{Graham1970}%
  \BibitemOpen
  \bibfield  {author} {\bibinfo {author} {\bibfnamefont {R.}~\bibnamefont
  {Graham}}\ and\ \bibinfo {author} {\bibfnamefont {H.}~\bibnamefont {Haken}},\
  }\href {\doibase 10.1007/BF01400474} {\bibfield  {journal} {\bibinfo
  {journal} {Zeitschrift fur Phys.}\ }\textbf {\bibinfo {volume} {237}},\
  \bibinfo {pages} {31} (\bibinfo {year} {1970})}\BibitemShut {NoStop}%
\bibitem [{\citenamefont {Pytte}\ and\ \citenamefont
  {Thomas}(1968)}]{Pytte1968}%
  \BibitemOpen
  \bibfield  {author} {\bibinfo {author} {\bibfnamefont {E.}~\bibnamefont
  {Pytte}}\ and\ \bibinfo {author} {\bibfnamefont {H.}~\bibnamefont {Thomas}},\
  }\href@noop {} {\bibfield  {journal} {\bibinfo  {journal} {Phys. Rev. Lett.}\
  }\textbf {\bibinfo {volume} {20}},\ \bibinfo {pages} {1167} (\bibinfo {year}
  {1968})}\BibitemShut {NoStop}%
\bibitem [{\citenamefont {Landauer}(1961)}]{Landauer1961}%
  \BibitemOpen
  \bibfield  {author} {\bibinfo {author} {\bibfnamefont {R.}~\bibnamefont
  {Landauer}},\ }\href@noop {} {\bibfield  {journal} {\bibinfo  {journal} {IBM
  J. Res. Dev.}\ }\textbf {\bibinfo {volume} {5}},\ \bibinfo {pages} {183}
  (\bibinfo {year} {1961})}\BibitemShut {NoStop}%
\bibitem [{\citenamefont {Landauer}(1971)}]{Landauer1971}%
  \BibitemOpen
  \bibfield  {author} {\bibinfo {author} {\bibfnamefont {R.}~\bibnamefont
  {Landauer}},\ }\href {\doibase 10.1080/00150197108243944} {\bibfield
  {journal} {\bibinfo  {journal} {Ferroelectrics}\ }\textbf {\bibinfo {volume}
  {2}},\ \bibinfo {pages} {47} (\bibinfo {year} {1971})}\BibitemShut {NoStop}%
\bibitem [{\citenamefont {Aspelmeyer}\ \emph {et~al.}(2014)\citenamefont
  {Aspelmeyer}, \citenamefont {Kippenberg},\ and\ \citenamefont
  {Marquardt}}]{Aspelmeyer2014}%
  \BibitemOpen
  \bibfield  {author} {\bibinfo {author} {\bibfnamefont {M.}~\bibnamefont
  {Aspelmeyer}}, \bibinfo {author} {\bibfnamefont {T.~J.}\ \bibnamefont
  {Kippenberg}}, \ and\ \bibinfo {author} {\bibfnamefont {F.}~\bibnamefont
  {Marquardt}},\ }\href {\doibase 10.1103/RevModPhys.86.1391} {\bibfield
  {journal} {\bibinfo  {journal} {Rev. Mod. Phys.}\ }\textbf {\bibinfo {volume}
  {86}} (\bibinfo {year} {2014}),\ 10.1103/RevModPhys.86.1391},\ \Eprint
  {http://arxiv.org/abs/0712.1618} {arXiv:0712.1618} \BibitemShut {NoStop}%
\bibitem [{\citenamefont {Xu}\ \emph {et~al.}(2017)\citenamefont {Xu},
  \citenamefont {Kemiktarak}, \citenamefont {Fan}, \citenamefont {Ragole},
  \citenamefont {Lawall},\ and\ \citenamefont {Taylor}}]{Xu2015b}%
  \BibitemOpen
  \bibfield  {author} {\bibinfo {author} {\bibfnamefont {H.}~\bibnamefont
  {Xu}}, \bibinfo {author} {\bibfnamefont {U.}~\bibnamefont {Kemiktarak}},
  \bibinfo {author} {\bibfnamefont {J.}~\bibnamefont {Fan}}, \bibinfo {author}
  {\bibfnamefont {S.}~\bibnamefont {Ragole}}, \bibinfo {author} {\bibfnamefont
  {J.}~\bibnamefont {Lawall}}, \ and\ \bibinfo {author} {\bibfnamefont {J.~M.}\
  \bibnamefont {Taylor}},\ }\href {\doibase 10.1038/ncomms14481} {\bibfield
  {journal} {\bibinfo  {journal} {Nat. Commun.}\ }\textbf {\bibinfo {volume}
  {8}},\ \bibinfo {pages} {14481} (\bibinfo {year} {2017})},\ \Eprint
  {http://arxiv.org/abs/1510.04971} {arXiv:1510.04971} \BibitemShut {NoStop}%
\bibitem [{\citenamefont {Haken}(1975)}]{Haken1975}%
  \BibitemOpen
  \bibfield  {author} {\bibinfo {author} {\bibfnamefont {H.}~\bibnamefont
  {Haken}},\ }\href {\doibase 10.1103/RevModPhys.47.67} {\bibfield  {journal}
  {\bibinfo  {journal} {Rev. Mod. Phys.}\ }\textbf {\bibinfo {volume} {47}},\
  \bibinfo {pages} {67} (\bibinfo {year} {1975})}\BibitemShut {NoStop}%
\bibitem [{\citenamefont {Jayich}\ \emph
  {et~al.}(2008{\natexlab{a}})\citenamefont {Jayich}, \citenamefont {Sankey},
  \citenamefont {Zwickl}, \citenamefont {Yang}, \citenamefont {Thompson},
  \citenamefont {Girvin}, \citenamefont {Clerk}, \citenamefont {Marquardt},\
  and\ \citenamefont {Harris}}]{Jayich2008a}%
  \BibitemOpen
  \bibfield  {author} {\bibinfo {author} {\bibfnamefont {a.~M.}\ \bibnamefont
  {Jayich}}, \bibinfo {author} {\bibfnamefont {J.~C.}\ \bibnamefont {Sankey}},
  \bibinfo {author} {\bibfnamefont {B.~M.}\ \bibnamefont {Zwickl}}, \bibinfo
  {author} {\bibfnamefont {C.}~\bibnamefont {Yang}}, \bibinfo {author}
  {\bibfnamefont {J.~D.}\ \bibnamefont {Thompson}}, \bibinfo {author}
  {\bibfnamefont {S.~M.}\ \bibnamefont {Girvin}}, \bibinfo {author}
  {\bibfnamefont {a.~a.}\ \bibnamefont {Clerk}}, \bibinfo {author}
  {\bibfnamefont {F.}~\bibnamefont {Marquardt}}, \ and\ \bibinfo {author}
  {\bibfnamefont {J.~G.~E.}\ \bibnamefont {Harris}},\ }\href {\doibase
  10.1088/1367-2630/10/9/095008} {\bibfield  {journal} {\bibinfo  {journal}
  {New J. Phys.}\ }\textbf {\bibinfo {volume} {10}} (\bibinfo {year}
  {2008}{\natexlab{a}}),\ 10.1088/1367-2630/10/9/095008},\ \Eprint
  {http://arxiv.org/abs/0805.3723} {arXiv:0805.3723} \BibitemShut {NoStop}%
\bibitem [{\citenamefont {Jayich}\ \emph
  {et~al.}(2008{\natexlab{b}})\citenamefont {Jayich}, \citenamefont {Sankey},
  \citenamefont {Zwickl}, \citenamefont {Yang}, \citenamefont {Thompson},
  \citenamefont {Girvin}, \citenamefont {Clerk}, \citenamefont {Marquardt},\
  and\ \citenamefont {Harris}}]{Jayich2008c}%
  \BibitemOpen
  \bibfield  {author} {\bibinfo {author} {\bibfnamefont {A.~M.}\ \bibnamefont
  {Jayich}}, \bibinfo {author} {\bibfnamefont {J.~C.}\ \bibnamefont {Sankey}},
  \bibinfo {author} {\bibfnamefont {B.~M.}\ \bibnamefont {Zwickl}}, \bibinfo
  {author} {\bibfnamefont {C.}~\bibnamefont {Yang}}, \bibinfo {author}
  {\bibfnamefont {J.~D.}\ \bibnamefont {Thompson}}, \bibinfo {author}
  {\bibfnamefont {S.~M.}\ \bibnamefont {Girvin}}, \bibinfo {author}
  {\bibfnamefont {A.~A.}\ \bibnamefont {Clerk}}, \bibinfo {author}
  {\bibfnamefont {F.}~\bibnamefont {Marquardt}}, \ and\ \bibinfo {author}
  {\bibfnamefont {J.~G.~E.}\ \bibnamefont {Harris}},\ }\href {\doibase
  10.1088/1367-2630/10/9/095008} {\bibfield  {journal} {\bibinfo  {journal}
  {New J. Phys.}\ }\textbf {\bibinfo {volume} {10}} (\bibinfo {year}
  {2008}{\natexlab{b}}),\ 10.1088/1367-2630/10/9/095008},\ \Eprint
  {http://arxiv.org/abs/0805.3723} {arXiv:0805.3723} \BibitemShut {NoStop}%
\bibitem [{\citenamefont {Sankey}\ \emph {et~al.}(2010)\citenamefont {Sankey},
  \citenamefont {Yang}, \citenamefont {Zwickl}, \citenamefont {Jayich},\ and\
  \citenamefont {Harris}}]{Sankey2010a}%
  \BibitemOpen
  \bibfield  {author} {\bibinfo {author} {\bibfnamefont {J.~C.}\ \bibnamefont
  {Sankey}}, \bibinfo {author} {\bibfnamefont {C.}~\bibnamefont {Yang}},
  \bibinfo {author} {\bibfnamefont {B.~M.}\ \bibnamefont {Zwickl}}, \bibinfo
  {author} {\bibfnamefont {A.~M.}\ \bibnamefont {Jayich}}, \ and\ \bibinfo
  {author} {\bibfnamefont {J.~G.~E.}\ \bibnamefont {Harris}},\ }\href {\doibase
  10.1038/nphys1707} {\bibfield  {journal} {\bibinfo  {journal} {Nat. Phys.}\
  }\textbf {\bibinfo {volume} {6}},\ \bibinfo {pages} {707} (\bibinfo {year}
  {2010})},\ \Eprint {http://arxiv.org/abs/1002.4158} {arXiv:1002.4158}
  \BibitemShut {NoStop}%
\bibitem [{\citenamefont {Xu}\ \emph {et~al.}(2016)\citenamefont {Xu},
  \citenamefont {Mason}, \citenamefont {Jiang},\ and\ \citenamefont
  {Harris}}]{Xu2016}%
  \BibitemOpen
  \bibfield  {author} {\bibinfo {author} {\bibfnamefont {H.}~\bibnamefont
  {Xu}}, \bibinfo {author} {\bibfnamefont {D.~J.}\ \bibnamefont {Mason}},
  \bibinfo {author} {\bibfnamefont {L.}~\bibnamefont {Jiang}}, \ and\ \bibinfo
  {author} {\bibfnamefont {J.~G.~E.}\ \bibnamefont {Harris}},\ }\href {\doibase
  10.1038/nature18604} {\bibfield  {journal} {\bibinfo  {journal} {Nature}\ ,\
  \bibinfo {pages} {1}} (\bibinfo {year} {2016})},\ \Eprint
  {http://arxiv.org/abs/1602.06881} {arXiv:1602.06881} \BibitemShut {NoStop}%
\bibitem [{\citenamefont {Chan}\ \emph {et~al.}(2011)\citenamefont {Chan},
  \citenamefont {Alegre}, \citenamefont {Safavi-Naeini}, \citenamefont {Hill},
  \citenamefont {Krause}, \citenamefont {Groeblacher}, \citenamefont
  {Aspelmeyer},\ and\ \citenamefont {Painter}}]{Chan2011}%
  \BibitemOpen
  \bibfield  {author} {\bibinfo {author} {\bibfnamefont {J.}~\bibnamefont
  {Chan}}, \bibinfo {author} {\bibfnamefont {T.~P.~M.}\ \bibnamefont {Alegre}},
  \bibinfo {author} {\bibfnamefont {A.~H.}\ \bibnamefont {Safavi-Naeini}},
  \bibinfo {author} {\bibfnamefont {J.~T.}\ \bibnamefont {Hill}}, \bibinfo
  {author} {\bibfnamefont {A.}~\bibnamefont {Krause}}, \bibinfo {author}
  {\bibfnamefont {S.}~\bibnamefont {Groeblacher}}, \bibinfo {author}
  {\bibfnamefont {M.}~\bibnamefont {Aspelmeyer}}, \ and\ \bibinfo {author}
  {\bibfnamefont {O.}~\bibnamefont {Painter}},\ }\href {\doibase
  10.1038/nature10461} {\bibfield  {journal} {\bibinfo  {journal} {Nature}\
  }\textbf {\bibinfo {volume} {478}},\ \bibinfo {pages} {18} (\bibinfo {year}
  {2011})},\ \Eprint {http://arxiv.org/abs/1106.3614} {arXiv:1106.3614}
  \BibitemShut {NoStop}%
\end{thebibliography}%

\end{document}